\documentclass[%
 aip,
 jmp,%
 amsmath,amssymb,
 reprint,%
]{revtex4-1}

\usepackage{graphicx}
\usepackage{dcolumn}
\usepackage{bm}

\begin{document}

\preprint{AIP/123-QED}

\title{On symmetries and conserved quantities in Nambu mechanics}

\author{M. Fecko}
\email{fecko@fmph.uniba.sk}
\affiliation{Department of Theoretical Physics, Comenius University, Bratislava, Slovakia}



\begin{abstract}
In Hamiltonian mechanics, a (continuous) symmetry leads to conserved quantity, which is a {\it function}
 on (extended) phase space.
 In Nambu mechanics, a straightforward consequence of symmetry is just a {\it relative integral invariant},
 a differential form which only upon integration over a cycle provides a conserved real number.
 The origin of the difference may be traced back to a shift in degrees of relevant forms present in equations of motion,
 or, alternatively, to a corresponding shift in degrees of relevant objects in {\it action integral}
 for Nambu mechanics.
\end{abstract}

\maketitle

\section{\label{sec:intro}Introduction}

According to seminal paper of Emmy Noether
(Ref.~\onlinecite{Noether1918}, see also Ref.~\onlinecite{Kosmann2011}),
there is a close correspondence between symmetries of action integral and conserved quantities
for the dynamics given by the action.

Recall briefly, how it works in \emph{Hamiltonian mechanics}. First, Hamilton equations
\begin{equation}
      {\dot q}^a = \frac{\partial H}{\partial p_a}
      \hskip 1cm
      {\dot p}_a = -\frac{\partial H}{\partial q^a}
      \label{hamiltoneq1}
\end{equation}
may be succinctly written as
\begin{equation}
      i_{\dot \gamma}d\sigma = 0
      \label{hamiltoneq2}
\end{equation}
(see Ref.~\onlinecite{Arnold1989, Fecko2006}), where
\begin{equation}
       \dot \gamma = {\dot q}^a\partial_{q^a}
                   + {\dot p}_a\partial_{p_a}
                   + \partial_t
       \label{hamiltoneq3}
\end{equation}
is the velocity vector to curve $\gamma$ (on \emph{extended} phase space) and
\begin{equation}
      \sigma = p_adq^a - Hdt
      \label{hamiltoneq4}
\end{equation}
is a distinguished 1-form, Poincar\'e-Cartan integral invariant (on extended phase space as well).
Then, the standard action integral reads (see Refs.~\onlinecite{Arnold1989, LandauLifshitz1995, Fecko2006})
\begin{equation}
      S[\gamma] = \int_\gamma \sigma
                = \int_{t_1}^{t_2}(p_a {\dot q}^a - H)dt
      \label{hamiltonaction}
\end{equation}
Conserved quantities extracted from (continuous) symmetries of the action are \emph{functions} on (extended) phase space
(e.g. ener\-gy from time translations, components of momentum from space translations,
components of angular momentum from rotations etc.).
Explicitly (for details, including a derivation, see Appendix \ref{apphamilton}),
if the symmetry is given by a vector field $\xi$ (on extended phase space),
the corresponding conserved quantity $f_\xi$ is given as
\begin{equation}
      f_\xi = i_\xi \sigma - \chi_\xi
      \label{hamiltonconserved}
\end{equation}

In 1973 Nambu proposed a modification of Hamiltonian mechanics (Ref.~\onlinecite{Nambu1973}).
In its basic version, phase space is 3-dimensional (so that extended Nambu phase space is 4-dimensional)
and equations of motion (Nambu equations) read
\begin{equation}
\dot x_i = \epsilon_{ijk} \frac{\partial H_1}{\partial x_j}
           \frac{\partial H_2}{\partial x_k}
           \hskip .8cm i=1,2,3
\label{nambueq1}
\end{equation}
or, in vector notation,
\begin{equation}
\dot {\mathbf r} = \boldsymbol \nabla H_1 \times \boldsymbol \nabla H_2
\label{nambueq2}
\end{equation}
Here, the two ``Hamiltonians" are, in general, functions of $x_1,x_2,x_3$ and $t$.

It turns out that construction of action integral for Nambu mechanics is a slightly delicate subject.
It was observed (Ref.~\onlinecite{Fecko1992, Takhtajan1994}) that Nambu equations
(\ref{nambueq1}) may be succinctly written as
\begin{equation}
      i_{\dot \gamma}d\hat \sigma = 0
      \label{nambueq3}
\end{equation}
too, where
\begin{equation}
       \dot \gamma = {\dot x^1}\partial_1
              + {\dot x^2}\partial_2
              + {\dot x^3}\partial_3
              + \partial_t
       \label{nambueq4}
\end{equation}
is the velocity vector to curve $\gamma$ on extended Nambu phase space.

Equation (\ref{nambueq3}) formally looks exactly like (\ref{hamiltoneq2}),
it has the structure of ``vortex-lines equation" (see Ref.~\onlinecite{Fecko2013}).
However, there is an important difference between the two equations,
in that the form $\hat \sigma$, the counterpart of the \emph{one}-form (\ref{hamiltoneq4}),
is a \emph{two}-form, now. Explicitly, it reads
\begin{equation}
       \hat \sigma := x^1dx^2 \wedge dx^3 - H_1dH_2 \wedge dt
       \label{nambueq5}
\end{equation}
At first sight, the difference might look innocent. Notice, however,
that it is no longer possible to write down action integral like (\ref{hamiltonaction}),
since there is no candidate for \emph{one}-form to be integrated
along the \emph{curve} $\gamma$. Instead, the \emph{two}-form $\hat \sigma$ is available.
Therefore, the only way to produce a number (the value of action) is to integrate $\hat \sigma$
over a~two-dimensional \emph{surface}.

In Ref.~\onlinecite{Fecko1992}, a possibility to associate a surface with a~\emph{single} trajectory
$\gamma$ is investigated. It leads to an action, proposed already before
in Ref.~\onlinecite{BayenFlato1975}. This is not very satis\-factory, since its extremals are curves
on which $\dot {\mathbf r}$
is just \emph{proportional}, not necessarily equal, to the r.h.s. of (\ref{nambueq2}).

A more interesting way of how to come to a surface is proposed in Ref.~\onlinecite{Takhtajan1994}.
There, the value of action integral is associated with an appropriate one-parameter \emph{family} of trajectories
rather than with a single trajectory.

\begin{figure}
\includegraphics[width=0.8\linewidth]{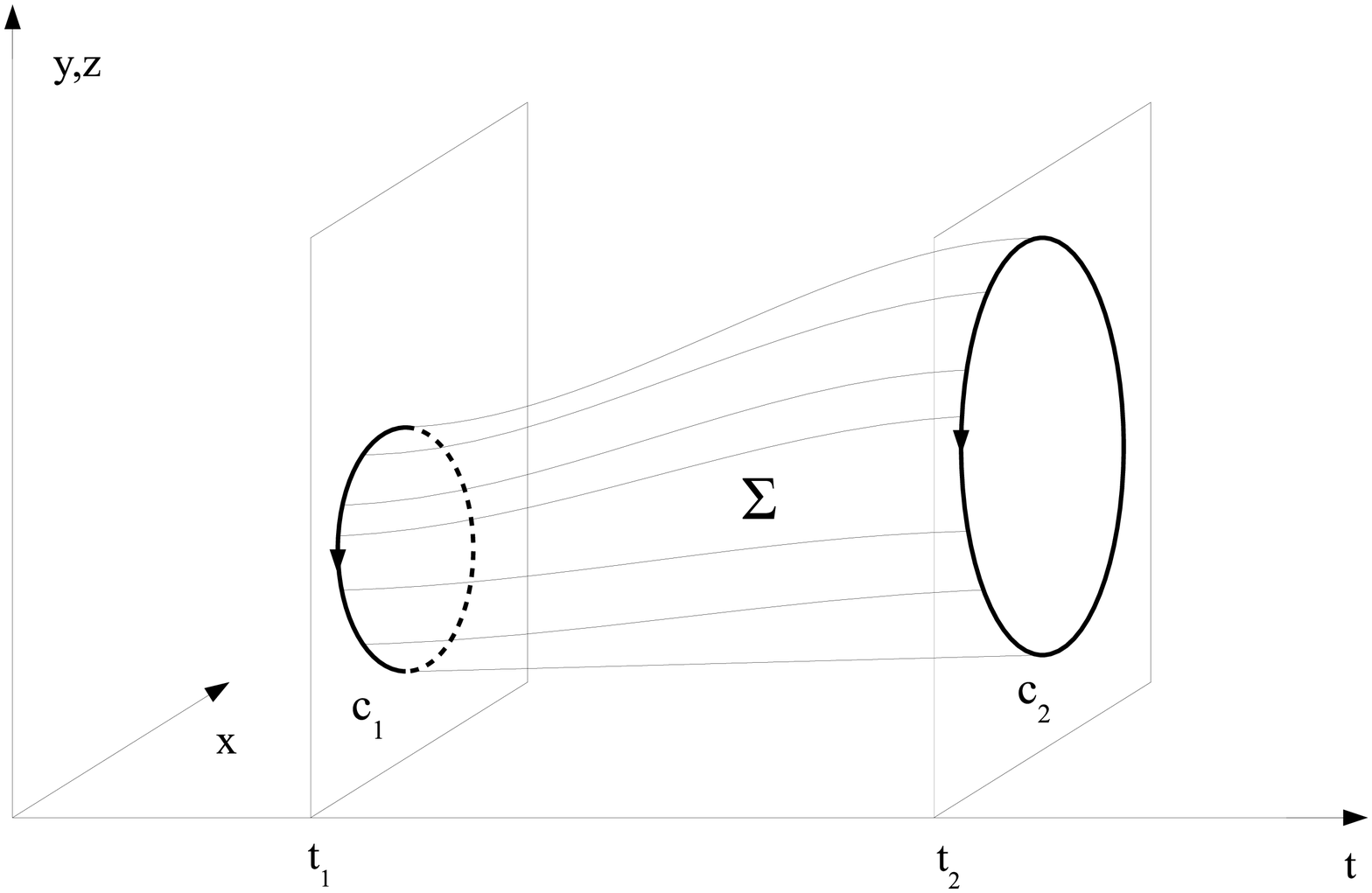}
\caption{\label{fig:epsart} A two-chain $\Sigma$ made up from a one-cycle $c_1$ using solutions
                            of Nambu equations.}
\end{figure}

Namely, consider the family constructed as follows:
Let, from each point $p$ of a \emph{one-cycle} (loop) $c_1$ at the time $t_1$,
emanate the solution $\gamma (t)$ of Nambu equations (\ref{nambueq3}),
fulfilling initial condition $\gamma (t_1)=p$.
At the time $t_2$, the points $\gamma (t_2)$ (for all $p\in c_1$) form a \emph{one-cycle} (loop) $c_2$ again
(the image of $c_1$ w.r.t. the Nambu flow for $t_2-t_1$)
and the points $\gamma(t)$,
for \emph{all} $t\in \langle t_1,t_2\rangle$ and \emph{all} $p\in c_1$, form a \emph{two-chain}
($2$-dimensional surface) $\Sigma$
made of solutions (see Fig.\ref{fig:epsart}; notice that $\partial \Sigma = c_1-c_2$).
The value of the action, assigned to the family, is defined as
\begin{equation}
      S[\Sigma] = \int_\Sigma \hat \sigma
      \label{nambuaction}
\end{equation}
One then easily verifies (see Appendix \ref{apptakhaction}) that the surface given by the family
of \emph{solutions} of Nambu equations is indeed an \emph{extremal} of the action integral
(\ref{nambuaction}).

\section{\label{sec:symmaction}Symmetries of the action}

Now, let us mimic in Nambu setting, i.e. using Takhtajan's action integral (\ref{nambuaction}),
the standard ``Hamiltonian" procedure for obtaining conserved quantities from symmetries
(see Appendix \ref{apphamilton}).

So we call, first, vector field $\xi$ a \emph{symmetry} if the action integral (\ref{nambuaction})
evaluated on $\Phi_\epsilon (\Sigma)$ (the flow $\Phi_\epsilon$ corresponds to $\xi$, here)
gives the same number as on $\Sigma$ itself
\begin{equation}
      S[\Phi_\epsilon \Sigma] =  S[\Sigma]
      \label{nambusymmetry1}
\end{equation}
(i.e. $\delta S=0$; there are \emph{no} restrictions on either time component of $\xi$
or the values of $\xi$ at the boundary $\partial \Sigma = c_1-c_2$ of $\Sigma$).
By direct computation of $\delta S$ (see Appendix \ref{apptakhaction}), we obtain
\begin{equation}
   \delta S
            = \epsilon \int_{\Sigma}i_{\xi}d\hat \sigma
            + \epsilon \oint_{\partial \Sigma} i_\xi \hat \sigma
            \label{deltaes4}
\end{equation}
Now, the first integral on the r.h.s. vanishes on the surface $\Sigma$ given by the family
of \emph{solutions} of Nambu equations
(by the argument mentioned in Appendix \ref{apptakhaction}:
$\dot \gamma$ is tangent to $\Sigma$ and, at the same time, it is annihilated by $d\hat \sigma$).
The second integral is over
$\partial \Sigma = c_1-c_2$ and the sum of both integrals on the r.h.s. of (\ref{deltaes4}) is to vanish.
So we get
\begin{equation}
       0 = \left(\oint_{c_1} - \oint_{c_2}\right)i_\xi \hat \sigma
       \label{deltaes5}
\end{equation}
or, equivalently,
\begin{equation}
       \oint_{c_1}i_\xi \hat \sigma = \oint_{c_2}i_\xi \hat \sigma
       \label{deltaes6}
\end{equation}
This is, however, nothing but a \emph{conservation law}: for \emph{solutions} of Nambu equations,
\begin{equation}
      f_\xi(t_1; c_1) = f_\xi(t_2;c_2)
      \label{nambusymmetry3}
\end{equation}
where $f_\xi$ is given by the \emph{integral}
\begin{equation}
      f(t_a;c_a) := \oint_{c_a}i_\xi \hat \sigma \hskip 1cm a=1,2
      \label{nambusymmetry4}
\end{equation}
In full analogy with the Hamiltonian case (see the text following (\ref{hamsymmetry4})),
a more general definition of symmetry is possible.
Rather than using differential version of (\ref{nambusymmetry1}), vanishing of the Lie derivative
\begin{equation}
      \mathcal L_\xi \hat \sigma = 0
      \label{condstrongnambu}
\end{equation}
we define \emph{symmetry of Nambu system} as a vector field $\xi$ obeying somewhat weaker condition,
\begin{equation}
      \mathcal L_\xi \hat \sigma = d\chi_\xi
      \label{condweaknambu1}
\end{equation}
(so, \emph{exactness} of the Lie derivative is enough). Or, by Cartan's formula,
\begin{equation}
      i_\xi d\hat \sigma = -d(i_\xi \hat \sigma - \chi_\xi)
      \label{condweaknambu2}
\end{equation}
Upon integration over the surface $\Sigma$ we get
\begin{equation}
        \int_{\Sigma}i_{\xi}d\hat \sigma
              = - \oint_{\partial \Sigma} (i_\xi \hat \sigma - \chi_\xi)
                \label{condweaknambuintegrated}
\end{equation}
Since the l.h.s. vanishes (on solutions), it holds
\begin{equation}
       \oint_{c_1}(i_\xi \hat \sigma - \chi_\xi) = \oint_{c_2}(i_\xi \hat \sigma - \chi_\xi)
       \label{nambuconservedrelat}
\end{equation}
So, we obtain the statement
\begin{equation}
      f_\xi(t_1; c_1) = f_\xi(t_2;c_2)
      \label{nambusymmetry5}
\end{equation}
where $f_\xi$ is given by the \emph{integral}
\begin{equation}
      f_\xi(t_a;c_a) := \oint_{c_a}(i_\xi \hat \sigma - \chi_\xi) \hskip 1cm a=1,2
      \label{nambusymmetry6}
\end{equation}
In more wordy formulation: Given a symmetry $\xi$ take, at time $t_1$, an arbitrary
one-cycle (loop) $c_1$. Compute the line integral
\begin{equation}
      \int_{c_1}(i_\xi \hat \sigma - \chi_\xi)
      \label{wordy1}
\end{equation}
Then, let each point of $c_1$ evolve by Nambu flow up to time $t_2$. You get another
one-cycle (loop), $c_2$. Compute, again, the line integral
\begin{equation}
      \int_{c_2}(i_\xi \hat \sigma - \chi_\xi)
      \label{wordy2}
\end{equation}
The statement is: You get the same number.

\section{\label{sec:intinvariants}Integral invariants}

What we obtained from symmetry $\xi$
is nothing but a~\emph{relative integral invariant} for Nambu dynamics.
In ge\-ne\-ral this is, by definition, a differential $p$-form $\alpha$ such that,
when integrated over a \emph{$p$-cycle}, it gives an invariant w.r.t. the dynami\-cal flow.
Put another way, if a dynami\-cal vector field $\Gamma$
generates the flow $\Phi_t$ (time evolution) and if $c_2$ is the $\Phi_t$-image of an \emph{arbitrary} \emph{$p$-cycle} $c_1$, then,
\begin{equation}
      \oint_{c_1}\alpha = \oint_{c_2}\alpha
      \label{intinvariantdef}
\end{equation}
(see, e.g., Refs.~\onlinecite{Cartan1922, Pars1965, LandauLifshitz1995, Arnold1989}).

In our case, the result (\ref{nambuconservedrelat}) may be regarded as the statement that
on Nambu extended phase space endowed with the dynamical vector field
$\Gamma$ defined by
\begin{equation}
      i_\Gamma d\hat \sigma = 0
      \label{defGamma}
\end{equation}
(see (\ref{nambueq3}))
we get, as a consequence of existence of a symmetry $\xi$, a relative integral invariant.
So (\ref{intinvariantdef}) holds for the \emph{one-form}
\begin{equation}
      \alpha = i_\xi \hat \sigma - \chi_\xi
      \label{ourintinvariant}
\end{equation}
Of course, as is always the case, our relative integral invariant then automatically yields
an \emph{absolute} integral invariant, integral of the \emph{exterior derivative} $d\alpha$ of $\alpha$
over \emph{any} two-chain (two-dimensional surface) $s$.
So, taking into account (\ref{condweaknambu2}),
\begin{equation}
       \int_{s_1}i_\xi d\hat \sigma =  \int_{s_2}i_\xi d\hat \sigma
       \label{nambuconservedabsol}
\end{equation}

\section{\label{sec:momentummap}An approach via momentum map}

\emph{Momentum map} is a powerful tool for studying connection between symmetries
and conserved quantities in Hamiltonian framework.

Standardly, it is a map from a~symplectic manifold (pha\-se space of a Hamiltonian system)
to the dual of the Lie algebra of the symmetry group acting on the phase space
(see, e.g., Refs.~\onlinecite{Arnold1989, Fecko2006, CrampinPirani1986}).

A~slight\-ly modified version enables one to treat momentum map as a map from \emph{extended} phase
space of the Hamiltonian system to the dual of the Lie algebra (see Appendix \ref{appmomentumextend}).

Here we mimic the derivation of momentum map within the framework of the \emph{extended Nambu} phase space.

So, consider an action of a Lie group $G$ on extended Nambu phase space. The action satisfies
$R_g^*d\hat \sigma = d\hat \sigma$ or, infinitesimally, $\mathcal L_{\xi_X}d\hat \sigma = 0$.
This can be rewritten as \emph{closedness} of \emph{two-form} $\beta_X$
\begin{equation}
      d\beta_X  = 0 \hskip 1.5cm
      \beta_X := i_{\xi_X} d\hat \sigma
      \label{betaX}
\end{equation}
(Notice that (\ref{betaX}) is a statement concerning \emph{complete Nambu system},
 rather than just the extended Nambu phase space alone, since $\hat \sigma$ contains
 both ``Hamiltonians" $H_1$ and $H_2$.)
Often $\beta_X$ happens to be \emph{exact}; then we get
\begin{equation}
      i_{\xi_X} d\hat \sigma = -dP_X
      \label{betaXexact}
\end{equation}
Here, $P_X$ is a \emph{one-form} (rather than a zero-form, i.e. a~func\-tion, as in Hamiltonian case).
We can achieve li\-ne\-a\-rity of $P_X$ in $X$ in a standard way (see Ref.~\onlinecite{Fecko2006}),
write $P_X=X^iP_i$ and introduce
\begin{equation}
      P:=P_iE^i
      \label{Nambumomentummap}
\end{equation}
What we obtained is a $\mathcal G^*$-valued \emph{one}-form on extended Nambu phase space
(rather than the corresponding $\mathcal G^*$-valued \emph{zero}-form = \emph{function}
known from Hamiltonian framework as \emph{the momentum map}).
Its $j$-th component one-form, $P_j$, is defined, according to (\ref{betaXexact}), by the equation
\begin{equation}
      i_{\xi_{E_j}} d\hat \sigma = -dP_j
      \label{betaXexactj}
\end{equation}
Comparison with (\ref{condweaknambu2}) and (\ref{nambuconservedrelat}) shows that in this way we get,
as a reward for finding a symmetry, \emph{the same} integral invariants as we found in
Section \ref{sec:symmaction},
\begin{equation}
      \oint_cP_j = \ \text{relative integral invariant}
      \label{jthintinvrel}
\end{equation}
($P_j$ equals $i_{\xi_{E_j}} \hat \sigma - \chi_{\xi_{E_j}}$ modulo additive \emph{closed} one-form,
 vanishing after integration over the cycle)
 and in Section \ref{sec:intinvariants},
\begin{equation}
      \int_si_{\xi_{E_j}} d\hat \sigma  = \ \text{absolute integral invariant}
      \label{jthintinvabs}
\end{equation}
(see (\ref{nambuconservedabsol}) and (\ref{betaXexactj})).

 So, to conclude, \emph{both} approaches (the one discussed in Section \ref{sec:symmaction} as well as the one discussed here)
 lead to \emph{the same picture} regarding the relation between symmetries and conserved quantities due to them:
 each symmetry provides us with a \emph{relative integral invariant} of the form (\ref{nambuconservedrelat})
 (and, consequently, with an \emph{absolute integral invariant} of the form (\ref{nambuconservedabsol})).

Let us remark that the question of momentum map in the context of Nambu mechanics was already addressed before
in Ref.~\onlinecite{PanditGangal1999}.
There, it was introduced as a $\mathcal G^* \times \mathcal G^*$ valued mapping (i.e. function) defined by the formula
\begin{equation}
      i_{\xi_X} d\hat \sigma = dP_{1X}\wedge dP_{2X}
      \label{panditgangal}
\end{equation}
where both components of the pair $(P_1,P_2)$ are defined by (\ref{mommapusual1}),
in which $M$ is \emph{Nambu} phase space, now. Notice, however, that the formula
(\ref{panditgangal}) is inconsistent, since the l.h.s. is linear in $X$ whereas
the r.h.s. is not.

\section{\label{sec:morenambuhamiltonians}The case of more Nambu Hamiltonians}

What we treated in detail was the ``basic" version of Nambu mechanics,
the situation, when the Nambu phase space is \emph{three}-dimensional and there are \emph{two}
Nambu ``Hamiltonians", $H_1$ and $H_2$.

Already in the original paper (Ref.~\onlinecite{Nambu1973}) Nambu pointed out that the idea may be
straightforwardly generalized to more dimensions,
$n$-dimensional (Nambu) phase space and $n-1$ Nambu ``Hamiltonians", $H_1,\dots H_{n-1}$.
(There are also other generalizations, see Refs.~\onlinecite{Nambu1973, Takhtajan1994}.)

And it is easily seen that all constructions discussed in this paper work equally well in the
$n$-dimensional version.
In particular, $\hat \sigma$ becomes $(n-1)$-form,
$c_1$ becomes $(n-2)$-cycle, $\Sigma$ is $(n-1)$-dimensional surface and so on.
(See Refs.~\onlinecite{Takhtajan1994, Fecko2013}.)
Conserved quantities are still integral invariants
(formally equally looking formulas (\ref{nambuconservedrelat}) and (\ref{nambuconservedabsol})
 hold, where $c_a$ are $(n-2)$-cycles and $s_a$ are $(n-1)$-chains).

\section{\label{sec:conclusions}Conclusions}

Both Hamiltonian and Nambu mechanics study motion of (formally speaking) \emph{points} in phase space
(or, by technical reasons, in extended phase space).
Therefore it is natural to expect conserved quantities to be \emph{functions}
on (perhaps extended) phase space.
Once we study a~par\-ti\-cular motion, we evaluate the function at the time $t_1$ at the point
where the motion begins, and then we profit from the fact that,
at the future points of the trajectory, the same value of the function is guaranteed by the conservation law.

In Hamiltonian mechanics the story really goes like this.
\emph{Functions} (like energy or various components of momentum) are often conserved and this fact
then makes life much more easy.

In Nambu mechanics, there are conserved \emph{functions} as well.
Already in the first paper on the subject (Ref.~\onlinecite{Nambu1973}), Nambu discusses, as a key example,
dynamical Euler equations for the motion of a free rigid rotator.
Here, the Nambu phase space is three-dimensional
 (actually, it is just a subsystem of a complete \emph{six}-dimensional \emph{Hamiltonian} system of equations;
  one should add \emph{kinematical} Euler equations to get the standard picture)
and both ener\-gy and square of the angular momentum are conserved.
Nambu shows that it is possible to choose these two (conserved) functions as the two ``Hamiltonians"
$H_1$ and $H_2$ in his approach.
In many papers, thereafter, authors write various systems of ordinary differential equations in Nambu mechanics form,
exactly to ``make explicit" conserved quantities (functions, namely $H_1$ and $H_2$).

However, the message of this paper is that \emph{these} conserved functions \emph{do not} directly follow
from symmetries, as is usual in Hamiltonian case.
In the case of symmetries, application of more or less standard machinery results,
because of a peculiar situation with the action integral
  (presence of a two-form rather than one-form, necessity of taking a \emph{family} of trajectories rather than a single trajectory),
in conserved quantities, which have the character of
\emph{integral invariants} rather then usual conserved functions.
Namely, the machinery leads to \emph{higher-degree} forms rather than usual zero-forms, that is, functions
(one-form for a relative invariant, and its exterior derivative, two-form, for the corresponding absolute invariant).
As a reward for finding a symmetry, the conserved \emph{number} is only obtained as \emph{integral} of the form over a one-cycle
(or two-chain for the absolute invariant).

In order to make the picture complete, let us note that also \emph{Liouville theorem} holds in Nambu me\-cha\-nics
(phase volume is conserved) irrespective of concrete Nambu ``Hamiltonians" (see Refs.~\onlinecite{Nambu1973, Fecko2013}).
This means that there is an integral invariant available, not related to symmetries, too.
(There is also whole series of well-known \emph{Poincar\'e-Cartan} integral invariants in Hamiltonian mechanics,
 with no relation to symmetries as well.)

So, there are altogether as many as \emph{three} kinds of conserved quantities in Nambu mechanics.
First, more common, evidently useful quantities (functions), which are, however, not related (at least in a clear way) to symmetries.
Second, more exotic quantities (integral invariants), which, on contrary, result from application of standard machinery on symmetries.
And third, the phase volume (integral invariant) which is not related to symmetries, again.


\appendix

\section{\label{apphamilton}Hamiltonian mechanics - extremals, symmetries and conserved quantities}

Here, in order to make comparison with Nambu mechanics easier, we recall briefly how standard
reasoning goes in \emph{Hamiltonian} mechanics.
So, our action integral is given by (\ref{hamiltonaction}).

On extended phase space, consider a vector field
with vanishing time component (otherwise yet arbitrary),
$W=W^a\partial_{q^a} + W_a\partial_{p_a} + 0.\partial_t$ (``varia\-tional" field).
Its infinitesimal flow $\Phi_\epsilon$ performs (``equal time") variations of curves
$\gamma \mapsto \gamma_\epsilon = \Phi_\epsilon (\gamma)$.
Then $S[\gamma] \mapsto S[\gamma_\epsilon]$, where
$$\begin{array} {rcl}
   S[\gamma_\epsilon] &=& \int_{\gamma_\epsilon}\sigma
         = \int_{\gamma}\Phi^*_\epsilon \sigma  =\int_{\gamma}(\hat 1 +\epsilon \mathcal L_W)\sigma          \\
         &=& S[\gamma] +\epsilon \int_{\gamma}i_Wd\sigma  +\epsilon \int_{\gamma}di_W\sigma                  \\
         &=& S[\gamma] +\epsilon \int_{\gamma}i_Wd\sigma  +\epsilon \int_{\partial \gamma}i_W\sigma          \\
  \end{array}
$$
So, we get for variation of action, $\delta S \equiv S[\gamma_\epsilon] - S[\gamma]$,
\begin{equation}
   \delta S
            = \epsilon \int_{t_1}^{t_2}\langle -i_{\dot \gamma}d\sigma , W \rangle dt
            + \epsilon (p_aW^a)|_{t_1}^{t_2}
            \label{deltaes3}
\end{equation}
This means that, within the class of curves with fixed $q^a$ at $t_1$ and $t_2$
(this corresponds to $W^a(\gamma(t_1))=0=W^a(\gamma(t_2))$),
extremals of the action ($\delta S=0$) coincide with
\emph{solutions} of equations of motion (\ref{hamiltoneq2}).

Now, vector field $\xi$ is a \emph{symmetry} if the action integral (\ref{hamiltonaction})
evaluated on $\Phi_\epsilon \circ \gamma$ (the flow $\Phi_\epsilon$ already corresponds to $\xi$, here)
gives the same number as on $\gamma$ itself
\begin{equation}
      S[\Phi_\epsilon \circ \gamma] =  S[\gamma]
      \label{hamsymmetry1}
\end{equation}
(i.e. $\delta S=0$; there are \emph{no} restrictions on either time component of $\xi$
or the values of $\xi$ at the ends of 
$\gamma$).
By the same direct computation as above, however, we obtain
\begin{equation}
   \delta S
            = \epsilon \int_{t_1}^{t_2}\langle -i_{\dot \gamma}d\sigma , \xi \rangle dt
            + \epsilon \int_{\partial \gamma} i_\xi \sigma
            \label{deltaes1}
\end{equation}
Therefore, combining  (\ref{hamsymmetry1}), (\ref{deltaes1}) and  (\ref{hamiltoneq2}) we see that
on \emph{solutions} of Hamilton equations one has
\begin{equation}
   0
            = \int_{\partial \gamma} i_\xi \sigma
\label{deltaes2}
\end{equation}
This is, however, nothing but a \emph{conservation law}: on solutions of Hamilton equations,
\begin{equation}
      f_\xi(\gamma(t_2)) = f_\xi(\gamma (t_1))
      \label{hamsymmetry3}
\end{equation}
for the function
\begin{equation}
      f_\xi:= i_{\xi} \sigma
      \label{hamsymmetry4}
\end{equation}
Actually, requiring (\ref{hamsymmetry1}) is too restrictive.
To see this notice that it is equivalent,
according to the first line of the computation in the beginning of this section, to
differential condition
\begin{equation}
      \mathcal L_\xi \sigma = 0
      \label{conditionstrong}
\end{equation}
It turns out, however, that a more general definition of symmetry may be useful, namely
as a vector field $\xi$ fulfilling just
\begin{equation}
      \mathcal L_\xi \sigma = d\chi_\xi
      \hskip 1cm \text{i.e.} \hskip 1cm
      i_\xi d\sigma = - d(i_\xi \sigma - \chi_\xi )
      \label{conditionweaker}
\end{equation}
(i.e. \emph{exactness} of the Lie derivative is enough for gaining a conserved quantity,
 its vanishing being too strong requirement).
Indeed, integrating (\ref{conditionweaker}) over $\gamma$ gives
\begin{equation}
      \int_{t_1}^{t_2}\langle i_{\dot \gamma}d\sigma , \xi \rangle dt
            = \int_{\partial \gamma} (i_\xi \sigma - \chi_\xi)
            \label{coditionintegrated}
\end{equation}
Since the l.h.s. vanishes (on solutions), we get conservation law (\ref{hamsymmetry3})
for more general function, namely
\begin{equation}
      f_\xi:= i_{\xi} \sigma - \chi_\xi
      \label{hamsymmetry5}
\end{equation}

\section{\label{apptakhaction}Takhtajan's action and its extremals}

Let us proceed to the \emph{Nambu} mechanics, now.
Consider Takhtajan's action integral (\ref{nambuaction}).
Infinitesimal flow $\Phi_\epsilon$ of variational field
$W=W^x\partial_x +W^y\partial_y +W^z\partial_z + 0.\partial_t$
performs (``equal time") variations of \emph{surfaces}
$\Sigma \mapsto \Sigma_\epsilon = \Phi_\epsilon (\Sigma)$.
Then $S[\Sigma] \mapsto S[{\Sigma}_\epsilon]$, where
$$\begin{array} {rcl}
   S[{\Sigma}_\epsilon]
   &=& \int_{{\Sigma}_\epsilon}\hat \sigma
    = \int_{\Sigma}\Phi^*_\epsilon \hat \sigma
    = \int_{\Sigma}(\hat 1 +\epsilon \mathcal L_W)\hat \sigma                                                       \\
   &=& S[\Sigma] +\epsilon \int_{\Sigma}i_Wd\hat \sigma  +\epsilon \int_{\Sigma}di_W\hat \sigma                  \\
   &=& S[\Sigma] +\epsilon \int_{\Sigma}i_Wd\hat \sigma  +\epsilon \int_{\partial \Sigma}i_W\hat \sigma          \\                                                        \\
  \end{array}
$$
Now, on $\Sigma$ made of \emph{solutions} of the equations of motion (\ref{nambueq3}),
integral $\int_{\Sigma}i_Wd\hat \sigma$ vanishes,
since we just sum terms proportional to $(d\hat \sigma)(W, \dot \gamma, u) = -(i_{\dot \gamma} d\hat \sigma)(W, u)=0$
($u$ is tangent to the surface $\Sigma$, linearly independent of $\dot \gamma$;
this is the counterpart of summing terms proportional to
 $(d\sigma)(W, \dot \gamma) = (-i_{\dot \gamma} d\sigma)(W)=0$ in the Hamiltonian case).
So we get for variation of action, $\delta S \equiv S[\Sigma_\epsilon] - S[\Sigma]$,
when computed on surface $\Sigma$ composed of \emph{solutions},
\begin{equation}
      \delta S  = \epsilon \left( \oint_{c_1} - \oint_{c_2} \right)x(W^ydz -W^zdy)
      \label{deltaes33}
\end{equation}
(since $i_W\hat \sigma = x(W^ydz -W^zdy) + (\dots)dt$ and $dt$ va\-ni\-shes on $c_1$ and $c_2$).
This means that surfaces made of solutions provide extremals ($\delta S=0$) in the class of
surfaces whose boundaries, 1-chains $c_1$ at $t_1$ and $c_2$ at $t_2$ respectively, have
fixed $y$ and $z$ values  (put another way, fixed projections onto the $yz$-plane).
This is because $W^y$ and $W^z$ should vanish at $t_1$ and $t_2$.
(There is no need to fix $x$ at the ends.
 This is a counterpart of the Hamiltonian freedom to move $p_a$ at the ends.)

\section{\label{appmomentumextend}Momentum map and extended phase space}

Most frequently, momentum map is associated with (certain) action of a Lie group $G$
on a phase space
(symplectic manifold $(M,\omega)$; see Refs.~\onlinecite{Arnold1989, CrampinPirani1986, Fecko2006}).
The action should preserve the symplectic form,
$R_g^*\omega = \omega$, so infinitesimally $\mathcal L_{\xi_X}\omega = 0$
(where $\xi_X$ is the generator of the action, $X\in \mathcal G$).
This can be rewritten as \emph{closedness} of $\alpha_X$
\begin{equation}
      d\alpha_X  = 0 \hskip 1.5cm
      \alpha_X := i_{\xi_X} \omega
      \label{alphaX}
\end{equation}
Often $\alpha_X$ happens to be \emph{exact}; then we get
\begin{equation}
      i_{\xi_X} \omega = -dP_X
      \hskip 1.5cm
      P_X:M\to \mathbb R
      \label{alphaXexact}
\end{equation}
Finally, since linearity of $P_X$ w.r.t. $X$ may always be achieved, we can introduce
momentum map as follows:
\begin{equation}
      P:M\to \mathcal G^*
      \hskip 1.5cm
      \langle P(m),X \rangle := P_X(m)
      \label{mommapusual1}
\end{equation}
Now, replace $\omega$ (on phase space) by $d\sigma$
(on extended phase space; $\sigma$ is given by (\ref{hamiltoneq4}))
\begin{equation}
      d\sigma = dp_a\wedge dq^a - dH \wedge dt
      \label{dsigma}
\end{equation}
So, consider an action of $G$ on \emph{extended} phase space $M\times \mathbb R$, such that
$R_g^*d\sigma = d\sigma$, so infinitesimally $\mathcal L_{\xi_X}d\sigma = 0$.
This can be rewritten as \emph{closedness} of $\beta_X$
\begin{equation}
      d\beta_X  = 0 \hskip 1.5cm
      \beta_X := i_{\xi_X} d\sigma
      \label{alphaXnew}
\end{equation}
(Notice that, unlike (\ref{alphaX}), which says something about the phase space alone,
 with no reference to particular Hamiltonian governing the dynamics,
 (\ref{alphaXnew}) is a statement concerning \emph{complete Hamiltonian} system,
  since $\sigma$ contains $H$.)
Often $\beta_X$ happens to be \emph{exact}; then we get
\begin{equation}
      i_{\xi_X} d\sigma = -dP_X
      \hskip 1cm
      P_X:M\times \mathbb R \to \mathbb R
      \label{alphaXexactnew}
\end{equation}
Finally, we can again introduce ``momentum map" as follows:
\begin{equation}
      P:M\times \mathbb R \to \mathcal G^*
      \hskip .7cm
      \langle P(m,t),X \rangle := P_X(m,t)
      \label{mommapusual2}
\end{equation}
(Unlike (\ref{alphaXexact}),
 we can \emph{not} read (\ref{alphaXexactnew}) as that the field $\xi_X$ is Hamiltonian;
 it lives on odd-dimensional manifold).
This function (i.e. all component ``ordinary" functions $P_i$, given by $P=P_iE^i$, $P_X=X^iP_i$) is \emph{conserved}.
Indeed, because of (\ref{alphaXexactnew}) and (\ref{hamiltoneq2}) we can write
$$\begin{array} {rcl}
   \dot P_X
   &\equiv& \dot \gamma P_X  = \langle dP_X,\dot \gamma \rangle = -(d\sigma)(\xi_X, \dot \gamma )  \\
   &=&  \langle i_{\dot \gamma}d\sigma,\xi_X \rangle =0                                            \\
  \end{array}
$$
so that
\begin{equation}
      \dot P_i =0
      \hskip 1cm
      i=1,\dots , \text{dim} \ \mathcal G
      \label{Pijenula}
\end{equation}
Comparison of (\ref{conditionweaker}) and (\ref{alphaXexactnew}) reveals, that the two ways
of obtaining conserved quantities from symmetry, discussed in Appendix \ref{apphamilton} and \ref{appmomentumextend}, respectively,
yield \emph{the same result}.
(Modulo, of course, an additive constant function.
 One has to \emph{fix} $X$ and call $\xi_X \equiv \xi$. Then $P_X$ from (\ref{alphaXexactnew})
 coincides with $f_\xi$ from (\ref{hamsymmetry5}).)

\nocite{*}
\bibliography{nambu_symm_revtex41}

\end{document}